\documentclass[conference]{IEEEtran}
\usepackage{multirow}
\usepackage{cite}

\usepackage[tight,footnotesize]{subfigure}
\usepackage{url}

\usepackage[pdftex]{graphicx}
\graphicspath{{./fig/}}
\DeclareGraphicsExtensions{.pdf,.jpeg,.png}

\usepackage[cmex10]{amsmath}
\usepackage{algorithmic}

\begin{document}

\title{Power and Energy-efficiency Roofline Model \\for GPUs}

\author{\IEEEauthorblockN{
Millad Ghane\IEEEauthorrefmark{1},
Jeff Larkin\IEEEauthorrefmark{2}, 
Larry Shi\IEEEauthorrefmark{1},
Sunita Chandrasekaran\IEEEauthorrefmark{3} and
Margaret S. Cheung\IEEEauthorrefmark{4}
}


\IEEEauthorblockA{\IEEEauthorrefmark{1}Department of Computer Science, University of Houston,
Houston, TX, USA\\ mghane@cs.uh.edu, wshi3@central.uh.edu}
\IEEEauthorblockA{\IEEEauthorrefmark{2}NVIDIA Corporation\\
jlarkin@nvidia.com}
\IEEEauthorblockA{\IEEEauthorrefmark{3}Computer \& Information Sciences, University of Delaware, 
Newark, DE, USA\\ schandra@udel.edu}
\IEEEauthorblockA{\IEEEauthorrefmark{4}Physics Department, University of Houston, Houston, TX, USA\\
\IEEEauthorrefmark{4}Center for Theoretical Biological Physics, Rice University, Houston, TX, USA\\ 
mscheung@central.uh.edu}

}

\maketitle
\begin{abstract}
Energy consumption has been a great deal of concern in recent years and developers need to take energy-efficiency into account when they design algorithms. Their design needs to be energy-efficient and low-power while it tries to achieve attainable performance provided by underlying hardware. However, different optimization techniques have different effects on power and energy-efficiency and a visual model would assist in the selection process.

In this paper, we extended the roofline model and provided a visual representation of optimization strategies for power consumption. Our model is composed of various ceilings regarding each strategy we included in our models. One roofline model for computational performance and one for memory performance is introduced. We assembled our models based on some optimization strategies for two widespread GPUs from NVIDIA: Geforce GTX 970 and Tesla K80.

\end{abstract}

\begin{IEEEkeywords}
Roofline Model; Performance; Power; Energy-efficiency; Low-power; Scientific Computing; 
\end{IEEEkeywords}
\section{Introduction}
Energy and power consumption have been playing a significant role in High Performance Computing (HPC) systems, especially with recent advances in hardware design \cite{fan2007power}. The current trend in HPC systems is towards building an energy-efficient design \cite{taylor2012dark}. Considering Tianhe-2 as the fastest supercomputer in 2015, power consumption of supercomputers are required to be 26 times more efficient with respect to their performance \cite{bailey2014adaptive} so that 20 MW limitation by DoE for power is not crossed. Furthermore, the current trend in designing datacenters leads to 400\% increase in the cost of cooling and power, and they are expected to continue to rise \cite{intelguys2008datacenter}. To that end, not only do hardware designers need to consider consumption of energy, application developers and algorithm designers are also required to deal with the consumption of energy and power as one of their design factors. The main goal of this paper is to demonstrate the relation among three pillars visually: performance, power consumption, and energy-efficiency of kernels in a single model in order to provide developers with insightful information.

Our model is inspired by the classic Roofline model \cite{williams2009roofline}. Such a model serves as a visual representation of operational intensity of a kernel against maximum attainable performance that the hardware can achieve. In their model, Williams et al. showed how peak performance and peak attainable memory bandwidth in their model relate to each other in a system. For small values of operational intensity, due to lack of the optimization of memory operations, a kernel is bounded by memory bandwidth of the system. 
For kernels that are not inherently memory-bound, improving locality turns them into compute-bound ones. 
In general, one can utilize roofline model to find bottlenecks in its kernels and help developers to find suitable technique to improve their performance.

The roofline model tries to incorporate both computation bounds and memory bounds into one model. Consequently, one looks into the model and identifies proper optimization techniques. However, the model does not provide insights on the power limitations of the system to the user. The assumption is that we have an unlimited source of energy and there is no concern on the amount of power and energy to be consumed, which is certainly incorrect for modern HPC centers. Therefore, a model that incorporates optimization techniques with respect to energy consumption would be useful to developers. 

Our contributions in this paper are as follows: 
\begin{itemize}
\item \textbf{Insightful visual representation:} Our model provides an insightful visual representation of power consumption with respect to energy efficiency in a system. Taking both power consumption and efficiency of computation and memory into consideration, we identify whether a kernel is power-bound (power-hungry) or compute-bound.

\item \textbf{Energy efficiency:} To characterize the efficiency of our architecture, we define energy efficiency as flops per Joule (J). We demonstrate the trade-off between energy-efficiency and power consumption in our model.

\item \textbf{Effects of optimization techniques on power and energy efficiency:} Through our models, developers could understand how optimization techniques would affect power and energy-efficiency and which technique has more impact on our kernel. 
\end{itemize}

\section{Construction of Roofline Models}
For simple von Neumann architecture \cite{choi2013roofline} the energy consumption is modeled as:

\begin{equation}
\label{EQ_E_init}
E = E_{flops} + E_{mem} + E_0
\end{equation}

\noindent where \(E_{flops}\) and \(E_{mem}\) stand for the total consumed energy for floating-point computations and memory operations, respectively. \(E_0\) is the constant energy required for our system to work. Our assumption is that \(E_0\) remains constant during the execution time. Table \ref{TB_symbols} describes the symbols on this paper. Considering Eq. \ref{EQ_E_init}, we rephrase it as following:

\begin{equation}
\label{EQ_E_rephrase}
E_C = W \epsilon_{flops} + Q \epsilon_{mem}
\end{equation}

\noindent where \(T\), \(W\), \(Q\) and \(E_C\) stand for the execution time, the number of floating-point operations, the number of memory operations, and the consumed energy for our kernel, respectively. 

Using simple mathematical model in Eq. \ref{EQ_E_rephrase} power consumption, \(P\), and reciprocal of energy efficiency, \(E_W = \frac{1}{EE}\), could be rewritten as following:

\begin{equation}
\label{EQ_E_T}
P = \frac{E_C}{t} = \frac{W}{t} \epsilon_{flop} + \frac{Q}{t} \epsilon_{mem} 
\end{equation}

\begin{equation}
\label{EQ_E_W}
E_W = \frac{E_C}{W} = \epsilon_{flop} + \frac{Q}{W} \epsilon_{mem}
\end{equation}

Combining Eq. \ref{EQ_E_T} and \ref{EQ_E_W} results in following linear equation between \(P\) and \(E_W\):

\begin{equation}
\label{EQ_Y_X}
P = \frac{W}{t} E_W
\end{equation}

On the other hand, under a specific optimization strategy the peak value for power consumption could be represented as \(P_{peak}\). Thus, a roofline model with relation between \(E_W\) and \(P\) could be defined as following:

\begin{equation}
\label{EQ_E_T_main}
P = min \{ \frac{E}{W} \times \pi , P_{peak} \}
\end{equation}

\noindent where \(\pi\) is the performance of system (defined as FLOP per second). 

Using the same approach, one can find similar relationship between energy per byte and power consumption as presented in Eq. \ref{EQ_P_Q_t_E_B}

\begin{equation}
\label{EQ_P_Q_t_E_B}
P = \frac{Q}{t} E_Q
\end{equation}

\noindent which \(\frac{Q}{t}\) and \(E_Q\) are the memory bandwidth (\(BW\)) and required energy to transfer all data to/from main memory on the device, respectively. Combining this equation with the peak value for power consumption leads to following relationship between \(P\) and \(\frac{E}{Q}\), which represents a roofline model for performance of memory subsystem. Equation \ref{P_EQ_BW_main} shows this roofline model.

\begin{equation}
\label{P_EQ_BW_main}
P = min \{ \frac{E}{Q} \times BW , P_{peak} \}
\end{equation}

\begin{table}[!t]
\renewcommand{\arraystretch}{1.3}
\caption{Symbols}
\label{TB_symbols}
\centering
\begin{tabular}{|c|l|}
\hline
\textbf{Symbol} & \textbf{Description} \\
\hline
\(W\) & \# of arithmetic operations (FLOP) \\
\hline
\(Q\) & \# of transferred bytes for memory operations \\
\hline
\(t\) & Total run time (Seconds) \\
\hline
\(E\) & Total consumed energy (Joules) \\
\hline
\(E_{flop}\) & Total energy of arithmetic operations (Joules) \\
\hline
\(E_{mem}\) & Total energy of transferred bytes (Joules) \\
\hline
\(E_{0}\) & Constant energy required for system to operate (Joules) \\
\hline
\(E_{C}\) & Required energy to run the kernel (Joules) \\
\hline
\(\epsilon_{flop}\) & Energy per arithmetic operation (Joules / FLOP) \\
\hline
\(\epsilon_{mem}\) & Energy per memory operation (Joules / Byte) \\
\hline
\(P_{peak}\) & Peak power consumption of system (Watts) \\
\hline
\(\pi\) & Computational performance of the kernel (FLOP / Second) \\
\hline
\(BW\) & Bandwidth (Bytes / Second) \\
\hline
\(EE_{comp}\) & Energy efficiency of computations (FLOP / Joule) \\
\hline
\(EE_{mem}\) & Energy efficiency of memory operations (FLOP / Joule) \\
\hline
\end{tabular}
\end{table}

A common metric to measure the energy-efficiency in HPC systems is performance-per-watts \cite{green500,scogland2013green500}, where performance is defined as "useful work" per second. Work, in scientific computing, is measured as the number of arithmetic operations, while in graph traversal algorithms, it could be defined as the number of traversed nodes in a graph \cite{choi2013roofline}. Therefore, in this paper, computational energy-efficiency is defined as following equation:

\begin{equation}
\label{EQ_EE_W}
EE_{comp} = \frac{Performance}{P} = \frac{\frac{W}{t}}{\frac{E}{t}} = \frac{W}{E}
\end{equation}

Mathematical elimination of \(t\) in Eq. \ref{EQ_EE_W} defines energy efficiency as the total number of floating-point operations per Joule. It shows that \(EE\) is in fact the reciprocal of X-axis in our model. Similar to this approach, energy efficiency of memory subsystem could be defined as memory bandwidth over consumed power as shown in Eq. \ref{EQ_EE_BW}.

\begin{equation}
\label{EQ_EE_BW}
EE_{mem} = \frac{Bandwidth}{P} = \frac{\frac{Q}{t}}{\frac{E}{t}} = \frac{Q}{E}
\end{equation}

Similar to Eq. \ref{EQ_EE_W}, Eq. \ref{EQ_EE_BW} would be the reciprocal of our X-axis for established roofline model of memory subsystem. Section \ref{SEC_ADD_CEILING} shows how Eq. \ref{EQ_E_T_main} and \ref{P_EQ_BW_main} will help us to build our models for two NVIDIA GPUs that we have considered in our paper for experimental analysis.

\section{Determining ceilings for Power and EE}
\label{SEC_ADD_CEILING}

The roofline models that we present in this paper are categorized into two groups: the first model tries to relate the energy-efficiency (EE) of computational subsystem with power and the second one relates EE of memory subsystem to power. Our models provide upper bound values for power and EE. This will help developers understand the optimizations that would result in less power consumption and/or better EE of the kernel. In other words, if a kernel were to be executed to collect its power and EE levels our model aims to determine the optimizations that should be implemented to make this kernel consume less power and/or be more energy efficient. Since there is a tradeoff between power consumption and EE it is a challenge to identify one unique technique that would lead to improving both levels simultaneously. 

Each optimization technique is represented as a ceiling in our model that designates the effect of applying the technique. The power and EE gap between two consecutive ceilings would show how much we will gain or lose by enabling associative technique.

\begin{figure}[!t]
\centering
\includegraphics[width=3.5in]{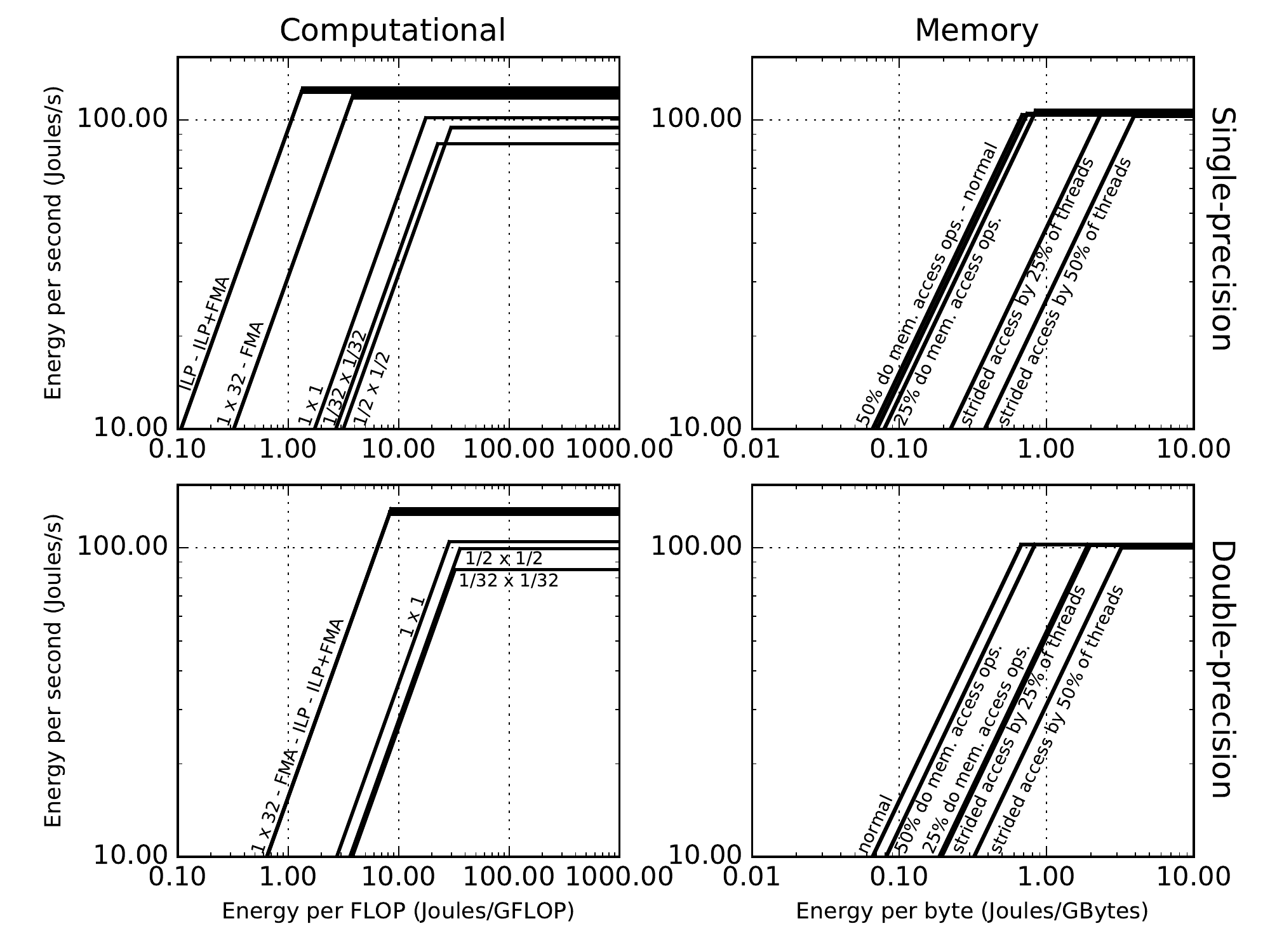}
\caption{Roofline Models for GTX 970 with different ceilings.}
\label{gtx970}
\end{figure}

Figure \ref{gtx970} depicts the computational and memory roofline models for single- and double-precision computations for NVIDIA Geforce GTX 970. To study the effect of computational performance, we implemented a simple reduction kernel in CUDA that computes dot product of two big arrays (each has 67,108,864 elements). At first, we changed number of threads and blocks. In the following figures, they are represented as the set of \(t \times b\) numbers, where t and b refer to the number of threads in a block and number of blocks, respectively. They are represented as a rational number w.r.t their peak value. Figure \ref{gtx970} demonstrates that increasing total number of threads results in a more energy-efficient kernel by losing a few Watts for single-precision computations. This statement is also correct when number of blocks is increased to an order of magnitude, otherwise, it does not help EE. The same trend is also noticeable for double-precision computations too. 

As the next step, we studied the effect of enabling fused multiply-add (FMA) operations. The FMA ceiling represents this optimization while number of threads and blocks are set to their peak values (1x32). Figure \ref{gtx970} shows enabling FMA has no significant effect on power and EE. 

The last step was to investigate the effect of instruction-level parallelism (ILP) through unrolling and maintaining partial sum of the main loop. The effect of such an optimization could easily be spotted for single-precision computations. In both cases of precisions, enabling ILP definitely enhances EE while only a few extra Watts is consumed. Above cases show the results for mutually independent studies of FMA and ILP techniques. However, we enabled them simultaneously and investigated their effect. Like our previous understandings, FMA operations did not affect performance and energy consumption significantly.

To study performance of memory subsystem, we used GPU-STREAM \cite{deakin2015gpu} as our benchmark application to measure the bandwidth of DRAM memory on the device. We modified the benchmark to support energy measurement by employing our Phoenix\footnote{https://github.com/milladgit/phoenix} library. Through imposing unoptimized modifications to our kernel we were able to investigate the effects of strided memory accesses. Furthermore, we also restricted memory accesses to a subset of threads to understand the effect of not exploiting parallelism in accessing arrays on the device in a uniform manner. Figure \ref{gtx970} depicts these effects on GTX 970. Strided memory accesses has significant effect on EE compared to limiting memory accesses to a small subset of total threads. If 50\% of threads access memory in a strided fashion, EE of our kernel will drop dramatically. This designates that if a kernel falls behind this ceiling, developers need to look into unwanted strided accesses to memory to find out the sources of losing EE in our kernels. In addition, since double-precision computations requires almost double bandwidth than single-precision ones, the effect of thread abandonment can be as severe as strided accesses. \textit{Normal} refers to uniform memory access among all threads in the system.

\begin{figure}[!t]
\centering
\includegraphics[width=3.5in]{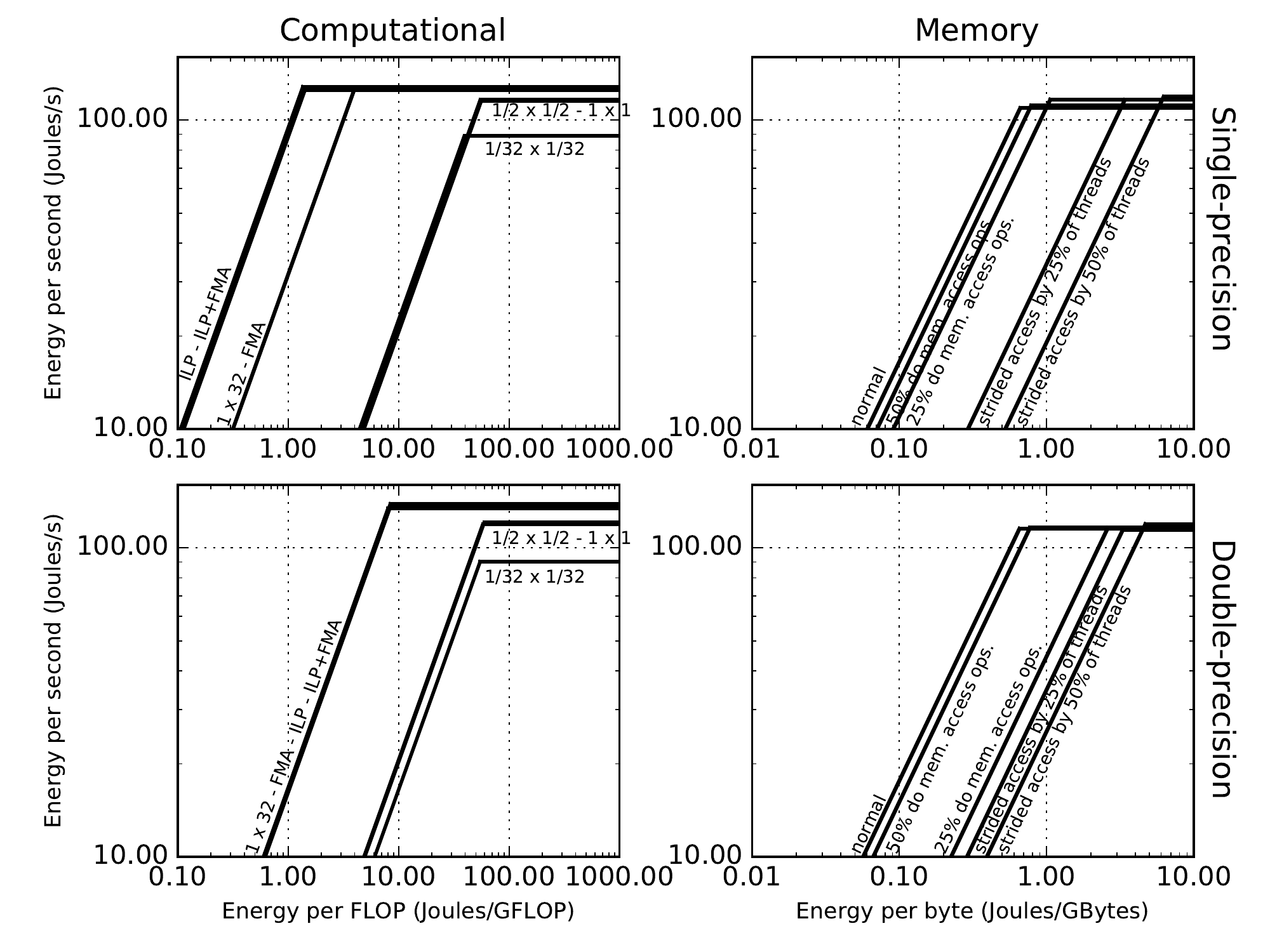}
\caption{Roofline Models for Tesla K80 with different ceilings.}
\label{k80}
\end{figure}

Figure \ref{k80} shows our models for NVIDIA Tesla K80. It depicts that the only way to gain efficiency in energy consumption is through implementing ILP or enabling FMA operations. Increasing number of threads and blocks leads to consuming more power while gaining no efficiency in energy consumption. When numbers of blocks are increased by orders of magnitude, we observed some EE, otherwise, like in GTX 970 increasing number of blocks in small steps does not help EE. All levels of parallelism should be enabled to achieve energy-efficiency in our designs. Memory ceilings of K80 follow a similar approach to GTX 970. Strided memory accesses on the device dramatically reduces our chances for EE. However, issuing memory accesses from a subset of threads (instead of uniform accesses to device memory) adversely affects EE when we are performing double-precision operations. 

\textbf{How does our model help developers?} 
We represent a kernel in terms of energy per flop and energy per byte representing the computational and memory performance of a given kernel. 
Position of these points with regards to the ceilings in our models will help developers identify relevant optimization techniques to improve power and EE of the kernel. It can be visually identified how much energy-efficient our system becomes by wasting a few Watts of power for each technique that we enable. 
For instance, if data point of the kernel falls behind ceilings of 1x1 for double-precision computations for K80, it indicates that either we need to implement ILP optimization technique in our kernel or increase the level of parallelism to its peak achievable value.

It should be noted that our models represent the relationship between power consumption and EE. Although our model does not identify optimization techniques for the developers (whereas roofline model does so), our model helps understand the influence of an optimization on power and energy efficiency. 
Nevertheless, one can confirm that the steepness of the line (slope) in our figures relate to the performance of computation (FLOP/s) and memory (BW); in both of our figures \ref{gtx970} and \ref{k80} the ceilings closer to the Y-axis represent better performance.

\section{Discussions on Related Work}

Needleess to say the Roofline model is a well established model by itself and there have been extensions proposed to the model in the past. Choi et. al. \cite{choi2013roofline} studied energy by using operational intensity as an independent variable to discuss the effect of energy and power on performance. Our model is inherently different to theirs as we are incorporating power consumption and energy efficiency into a single roofline model given known performance and memory bandwidth. Choi's model depicts how the power level relates to operational intensity of the kernel. However, it does not consider power and energy-efficiency in one model. 

In \cite{hong2010integrated}, Hong and Kim present a \textit{prediction} model for GPUs where the optimal number of active processors for any given program is predicted and that increasing number of cores for memory-bound applications does not improve computational performance. The GPUWattch model \cite{leng2013gpuwattch} presents a prediction model that accurately follows the power consumption footprint overtime. They also investigated the effect of DVFS using their model on GPUs. In our model, we do not predict a kernel's power consumption but we explore the relationship of power consumption w.r.t energy-efficiency of a kernel. 

Caparr{\'o}s and P{\"u}schel \cite{cabezas2014extending} proposed to evaluate performance by extracting the rooflines with the aid of  cycle-by-cycle analysis of the schedule that identifies the bottlenecks of underlying architecture. Their goal is to introduce additional rooflines to the model through a set of detailed architectural abstractions. They developed a mathematical model for performance based on a set of performance-relevant parameters from a modern processor by exploiting the extracted directed acyclic graph (DAG) of the computation. Ofenbeck et al. \cite{ofenbeck2014applying} produced a model through measuring a set of relative performance counters, like number of SSE and AVX instructions. We followed the same approach and calculate the number of floating-point operations as the amount of work to be done.

Illic et al. \cite{ilic2014cache} extended the original roofline model by proposing to measure the bandwidth observed from different cache levels instead of the whole DRAM memory in a multilevel cache hierarchy system. Therefore, for each cache level, they proposed a ceiling based on the bandwidth of that cache level. This approach leads to a robust model independent of the size of an input data. However, the original roofline model is dependent on the size of the input data. In a subsequent study, they incorporated power and energy consumption in their model \cite{ilic2016beyond} by proposing a mathematical formula for consumed energy and power at various levels of cache hierarchy. As shown in our results, although we observe that the data size may not influence the energy efficiency, it seem to certainly affect the power consumption.

In \cite{li2015transit} and \cite{li2016x}, authors modeled the system as an interactive queuing network and presented a visualized model similar to the roofline model. Through a small set of parameters from architecture and application one can build the model and study the effect of different levels of parallelism on their application. Nevertheless, their model did not include power consumption and energy-efficiency and they merely focused on performance. 

There have been other efforts on roofline models for GPUs too. Nugteren et. al. \cite{nugteren2014roofline} investigated the effects of enabling DVFS on performance and represent its effects in the roofline model. Jia et. al. \cite{Jia2012} extracted roofline model for GPUs and demonstrated their generated model for NVIDIA C2050 and AMD HD5850. They introduced a set of common ceilings for both architectures.

\section{Conclusion and future work}

In this paper, we introduced two roofline models, inspired from the original roofline model. We aim to provide insightful data to application developers and algorithm designers on energy-efficiency and power consumption of a kernel. We developed a mathematical model of energy consumption and extended the traditional roofline model with both energy consumption and power consumption. Our roofline models consists of a set of ceilings that represent optimization techniques. Through these ceilings one can visually realize the effect of applying the techniques on power and energy-efficiency and accordingly achieve a low-power design. Currently our work in progress includes applying our model on real world- scientific molecular dynamics codes. 


\IEEEpeerreviewmaketitle
\bibliographystyle{IEEEtran}
\bibliography{references}

\begin{thebibliography}{10}
\providecommand{\url}[1]{#1}
\csname url@samestyle\endcsname
\providecommand{\newblock}{\relax}
\providecommand{\bibinfo}[2]{#2}
\providecommand{\BIBentrySTDinterwordspacing}{\spaceskip=0pt\relax}
\providecommand{\BIBentryALTinterwordstretchfactor}{4}
\providecommand{\BIBentryALTinterwordspacing}{\spaceskip=\fontdimen2\font plus
\BIBentryALTinterwordstretchfactor\fontdimen3\font minus
  \fontdimen4\font\relax}
\providecommand{\BIBforeignlanguage}[2]{{%
\expandafter\ifx\csname l@#1\endcsname\relax
\typeout{** WARNING: IEEEtran.bst: No hyphenation pattern has been}%
\typeout{** loaded for the language `#1'. Using the pattern for}%
\typeout{** the default language instead.}%
\else
\language=\csname l@#1\endcsname
\fi
#2}}
\providecommand{\BIBdecl}{\relax}
\BIBdecl

\bibitem{fan2007power}
X.~Fan, W.-D. Weber, and L.~A. Barroso, ``Power provisioning for a
  warehouse-sized computer,'' in \emph{ACM SIGARCH Computer Architecture News},
  vol.~35, no.~2.\hskip 1em plus 0.5em minus 0.4em\relax ACM, 2007, pp. 13--23.

\bibitem{taylor2012dark}
M.~B. Taylor, ``Is dark silicon useful?: harnessing the four horsemen of the
  coming dark silicon apocalypse,'' in \emph{Proceedings of the 49th Annual
  Design Automation Conference}.\hskip 1em plus 0.5em minus 0.4em\relax ACM,
  2012, pp. 1131--1136.

\bibitem{bailey2014adaptive}
P.~E. Bailey, D.~K. Lowenthal, V.~Ravi, B.~Rountree, M.~Schulz, and B.~R.
  de~Supinski, ``Adaptive configuration selection for power-constrained
  heterogeneous systems,'' in \emph{International Conference on Parallel
  Processing}, vol.~43, 2014.

\bibitem{intelguys2008datacenter}
D.~Filani, J.~He, S.~Gao, M.~Rajappa, A.~Kumar, P.~Shah, and R.~Nagappan,
  ``Dynamic data center power management: Trends, issues, and solutions.''
  \emph{Intel Technology Journal}, vol.~12, no.~1, pp. 59 -- 67, 2008.

\bibitem{williams2009roofline}
S.~Williams, A.~Waterman, and D.~Patterson, ``Roofline: an insightful visual
  performance model for multicore architectures,'' \emph{Communications of the
  ACM}, vol.~52, no.~4, pp. 65--76, 2009.

\bibitem{choi2013roofline}
J.~W. Choi, D.~Bedard, R.~Fowler, and R.~Vuduc, ``A roofline model of energy,''
  in \emph{Parallel \& Distributed Processing (IPDPS), 2013 IEEE 27th
  International Symposium on}.\hskip 1em plus 0.5em minus 0.4em\relax IEEE,
  2013, pp. 661--672.

\bibitem{green500}
``Green500,'' \url{http://www.green500.org/}, accessed: 2016-10-20.

\bibitem{scogland2013green500}
T.~Scogland, B.~Subramaniam, and W.-c. Feng, ``The green500 list: escapades to
  exascale,'' \emph{Computer Science-Research and Development}, vol.~28, no.
  2-3, pp. 109--117, 2013.

\bibitem{deakin2015gpu}
T.~Deakin and S.~McIntosh-Smith, ``Gpu-stream: benchmarking the achievable
  memory bandwidth of graphics processing units.''

\bibitem{hong2010integrated}
S.~Hong and H.~Kim, ``An integrated gpu power and performance model,'' in
  \emph{ACM SIGARCH Computer Architecture News}, vol.~38, no.~3.\hskip 1em plus
  0.5em minus 0.4em\relax ACM, 2010, pp. 280--289.

\bibitem{leng2013gpuwattch}
J.~Leng, T.~Hetherington, A.~ElTantawy, S.~Gilani, N.~S. Kim, T.~M. Aamodt, and
  V.~J. Reddi, ``Gpuwattch: enabling energy optimizations in gpgpus,'' in
  \emph{ACM SIGARCH Computer Architecture News}, vol.~41, no.~3.\hskip 1em plus
  0.5em minus 0.4em\relax ACM, 2013, pp. 487--498.

\bibitem{cabezas2014extending}
V.~C. Cabezas and M.~P{\"u}schel, ``Extending the roofline model: Bottleneck
  analysis with microarchitectural constraints,'' in \emph{Workload
  Characterization (IISWC), 2014 IEEE International Symposium on}.\hskip 1em
  plus 0.5em minus 0.4em\relax IEEE, 2014, pp. 222--231.

\bibitem{ofenbeck2014applying}
G.~Ofenbeck, R.~Steinmann, V.~Caparros, D.~G. Spampinato, and M.~P{\"u}schel,
  ``Applying the roofline model,'' in \emph{Performance Analysis of Systems and
  Software (ISPASS), 2014 IEEE International Symposium on}.\hskip 1em plus
  0.5em minus 0.4em\relax IEEE, 2014, pp. 76--85.

\bibitem{ilic2014cache}
A.~Ilic, F.~Pratas, and L.~Sousa, ``Cache-aware roofline model: Upgrading the
  loft,'' \emph{IEEE Computer Architecture Letters}, vol.~13, no.~1, pp.
  21--24, 2014.

\bibitem{ilic2016beyond}
------, ``Beyond the roofline: Cache-aware power and energy-efficiency modeling
  for multi-cores,'' \emph{IEEE Transactions on Computers}, vol.~PP, no.~99,
  pp. 1--1, 2016.

\bibitem{li2015transit}
A.~Li, Y.~Tay, A.~Kumar, and H.~Corporaal, ``Transit: A visual analytical model
  for multithreaded machines,'' in \emph{Proceedings of the 24th International
  Symposium on High-Performance Parallel and Distributed Computing}.\hskip 1em
  plus 0.5em minus 0.4em\relax ACM, 2015, pp. 101--106.

\bibitem{li2016x}
A.~Li, S.~L. Song, E.~Brugel, A.~Kumar, D.~Chavarr{\'\i}a-Miranda, and
  H.~Corporaal, ``X: A comprehensive analytic model for parallel machines,'' in
  \emph{Parallel and Distributed Processing Symposium, 2016 IEEE
  International}.\hskip 1em plus 0.5em minus 0.4em\relax IEEE, 2016, pp.
  242--252.

\bibitem{nugteren2014roofline}
C.~Nugteren, G.-J. van~den Braak, and H.~Corporaal, ``Roofline-aware dvfs for
  gpus,'' in \emph{Proceedings of International Workshop on Adaptive
  Self-tuning Computing Systems}.\hskip 1em plus 0.5em minus 0.4em\relax ACM,
  2014, p.~8.

\bibitem{Jia2012}
\BIBentryALTinterwordspacing
H.~Jia, Y.~Zhang, G.~Long, J.~Xu, S.~Yan, and Y.~Li, \emph{GPURoofline: A Model
  for Guiding Performance Optimizations on GPUs}.\hskip 1em plus 0.5em minus
  0.4em\relax Berlin, Heidelberg: Springer Berlin Heidelberg, 2012, pp.
  920--932. [Online]. Available:
  \url{http://dx.doi.org/10.1007/978-3-642-32820-6_90}
\BIBentrySTDinterwordspacing

\end{thebibliography}

\end{document}